\documentclass[twocolumn,showpacs,preprintnumbers,amsmath,amssymb]{revtex4}

\usepackage{graphicx}
\usepackage{dcolumn}
\usepackage{bm}

\begin{document}

\title{The Ground State of a Large Number of Particles on a Frozen Topography}


\author{A. Travesset}
\affiliation{ Physics Department, Iowa
State University and Ames Lab.\\
Ames, IA, 50011, USA\\
}

\begin{abstract}
Problems consisting in finding the ground state of particles interacting with a given potential
constrained to move on a particular geometry are surprisingly difficult. Explicit solutions
have been found for small number of particles by the use of numerical methods in some particular
cases such as particles on the sphere and to a much lesser extent on a torus. In this paper
we propose a general solution to the problem in the opposite limit of very large number of particles $M$
by expressing the energy as an expansion in $M$ whose coefficients can be minimized by a
geometrical ansatz. The solution is remarkably universal with respect to the geometry and the interaction
potential. Explicit solutions for the sphere and the torus are provided. The paper concludes
with several predictions that could be verified by further theoretical or numerical work.

\end{abstract} \vfill
\maketitle
\section{Introduction}

Problems related to determining optimal particle distributions
under constraints are ubiquitous in the traditional
sciences and have been under intense scrutiny in the mathematical
community \cite{ConSl:92,SaKui:97}. Two of the many interesting examples
are the determination of ground states of $M$ particles constrained to move
on the sphere and interacting with a Coulomb potential, so called
Thomson problem \cite{Thoms:04}, whose more direct representation
are classical electrons on helium bubbles \cite{AlLei:92} and the
crystallization of particles on spheres, relevant for understanding the
structure of PMMA beads on oil/water droplets \cite{Science:03}.
Similar problems on more general geometries such as the torus or
negative curvature surfaces are also of great experimental and
theoretical interest \cite{BNTr:00,HarSa:04,HarSa:05}.

Theoretical investigations are surprisingly difficult. Extensive
numerical results obtained in the Thomson problem,
\cite{DTMHo:96,AWRTS:97,ErHoc:k97}, for
example, show that the number of metastable states grow very fast with the
total number of particles, preventing a
numerical solution to the problem even for a number of particles of a few hundred
particles. For problems on the sphere, a few rigorous analytical results and
conjectures on the energy of the ground state for large number of particles exist \cite{SaKui:97,HarSa:05},
but a description of the structure of these ground states, including practical tools on
how to find them as well as its generalization to any given arbitrary
geometry remains a completely open problem.

Recently, it has been shown that elasticity theory
\cite{BNTr:00,BoTra:01b,TrBCN:02,Trav:03} (see also \cite{DMoo:95,DPMoo:97})
provides a powerful framework to discuss best particle configurations
on spheres, which can then be
easily generalized to deal with any arbitrary geometry. Building
on these results, we propose a general solution for the structure
of ground states on arbitrary
geometries in the limit of large number of particles, and we construct the
explicit solution for the case of a sphere and a torus. The
solution is universal, in the sense that it applies for short-ranged
potentials and, in some situations, for long-ranged potentials as well.

\begin{figure}[htb]
\begin{center}
\includegraphics[width=2.5 in]{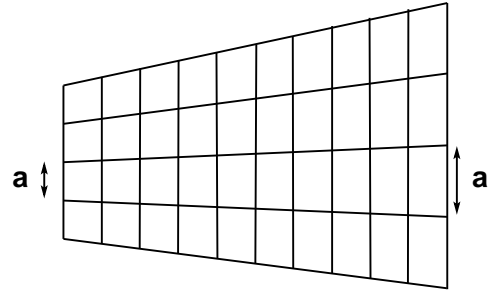}
\end{center}
\caption{Typical distortions due to geometric constraints on
a perfect square lattice. If an additional row of atoms somewhere in
the middle of the crystal were added, variations of the
lattice constant would become very small.}\label{fig__Argument}
\end{figure}

In this paper, we will argue that the problem of finding the
ground state of particles under constraints is equivalent to finding the particle distribution that
is closer to a perfectly equilateral triangulation \cite{com}(the triangulation constructed from the actual
distribution of particles via its Delaunay/Voronoi construction). The constraints that we
consider in this paper are either geometrical (particles are constrained on spheres, torus, etc..) or
topological (particles on a disk with the constraint that the total disclination charge is non-zero). If
the particles are on a plane, the absolute ground state is a triangular lattice. The question is how the
geometrical or topological constraints will modify this lattice.

The effect of the constraints is to induce spatial variations of the
lattice constant, as it is sketched in figure~\ref{fig__Argument} for a square lattice. This
is the same situation that was first addressed by Frank\cite{Frank:52} in the context
of crystal growth, where he showed that in order to minimize the strains induced by the variable
spacing it was necessary to add additional rows of atoms, that is, dislocations,
which correct the spatial variations in the lattice constant. Another critical result that we
use in this paper is that the energy of particles interacting with a given potential can be regarded as
an expression for large number of particles $M$ \cite{TrBCN:02,BNT05}, whose leading coefficient contains
a term, the $C-$function, which is always positive and encodes the dependence on both geometry and topological
defects. In this paper we will show how by adding dislocations it is possible to construct particle densities such that
the $C-$coefficient becomes identically zero
at leading order in $M$ (reaches its minimum value) thus providing explicit ground states
for very large number of particles.

Although our arguments will be
restricted to crystallization driven by energy, we believe that the geometric arguments
used to construct ground states are general and apply to entropic driven crystallization
such as for hard sphere potentials, since the entropic crystallization amounts to
maximizing the area of the unit cell for a given packing fraction, and for larger packing
fraction this leads to triangular lattices.

The organization of the paper is as follows. In
section~\ref{SECT_Potential} we review several results regarding
the expansion of the energy for large number of particles. In
section~\ref{SECT_ansatz} we implement the solution outlined in
fig.~\ref{fig__Argument} for a general triangular wedge, and it is
shown in section~\ref{SECT_Min} that the ansatz does minimize the
energy. Explicit results for the sphere and the torus are provided
in section~\ref{SECT_patch}. We end with a summary of predictions
that follow from this paper as well as with some conclusions in
section~\ref{SECT_conclusions}.

\section{Energy in the limit of large number of
particles}\label{SECT_Potential}

The energy of $M$ particles interacting with a potential
$V(\vec{r})$ is given by
\begin{equation}\label{energy_exact}
    E(M)=\frac{1}{2} \sum_{i,j} V(\vec{r}(i)-\vec{r}(j))
\end{equation}
where the sum runs over all $M$ particles at positions
$\vec{r}(i),(i=1..M)$. For definiteness, we discuss the concrete potential
\begin{equation}\label{concrete_pot}
    V(\vec{r})=\frac{e^2}{|\vec{r}|^s} ,
\end{equation}
but with minor modifications, the results can be made
general to include any repulsive potentials. If the particles are
arranged in a configuration close to a triangular lattice, we
write $\vec{r}(i)=\vec{R}+\vec{u}(i)+\vec{h}(i)$, where $\vec{R}(i)=a(n
e_1+me_2)$ define the vertices of a triangular lattice of lattice
constant $a$ and primitive vectors $e_1,e_2$, and $\vec{u},\vec{h}$ are
small quantities, in the sense that $\frac{|\vec{u}|}{a} <<
1$. The quantity $u$ represents distortions tangent to the plane where $h$
in the perpendicular direction. The energy Eq.~\ref{energy_exact} is
\begin{eqnarray}\label{Eq_elast}
E(M)&=& \frac{M}{2} \sum_{n,m}
    \frac{e^2}{|\vec{R}(n,m)|^s}
    +\frac{e^2}{2}\sum_{i,j}\Pi_{\alpha\beta}(i,j)
    u_{\alpha}(i)u_{\beta}(j) \nonumber\\
    &+& \frac{e^2}{2}\sum_{i,j}\Pi^0(i,j)h(i)h(j)
    +{\cal O}(u^3,u^2h,h^3))
\end{eqnarray}
where many higher order terms are neglected because of the assumed
smallness of the displacements $\vec{u},h$. On a more rigorous basis,
this step also requires the potential to be
short ranged (or $s>2$), thus excluding the Coulomb potential, although
for some geometries such as the sphere, we expect Eq.~\ref{Eq_elast} to hold as well, as it
will be shown. In the above expression, the contributions related to the geometry and topological defects are
entirely determined by the terms after the first.

We now review how the energy can be regarded as an expansion in large
number of particles. Detailed derivations have already been presented
somewhere else \cite{TrBCN:02,BNT05}, so
we just recall the main results. To avoid excessive generality and keep the
derivation simple, the results will be illustrated for the sphere. The
first term in Eq.~\ref{Eq_elast} gives the following explicit expression (for $0<s<2$)
\begin{eqnarray}\label{energy_planar}
\frac{M}{2} \sum_{n,m}
    \frac{e^2}{|\vec{R}(n,m)|^s}&=&\frac{e^2}{R^s}\left(\frac{M^2}{2^{s}(2-s)}+\right.\\\nonumber
      &+&\left.\frac{\theta(s)}{2(4\pi)^{s/2}}M^{1+s/2}+{\cal O}(M^{s/2}) \right)
\end{eqnarray}
where $R$ is the sphere radius. The only modification for $s>2$,
is that the $M^2$ term, which arises from the long range nature of
the potential, is absent. The actual values for the function
$\theta(s)$ maybe found in \cite{BNT05,HarSa:04}. The second contribution in
Eq.~\ref{Eq_elast} is evaluated by retaining the leading term in an expansion in
derivatives, leading to the familiar expression from elasticity theory
\begin{eqnarray}\label{energy_manifold}
&&\frac{e^2}{2}\sum_{i,j}\Pi_{\alpha\beta}(i,j)
    u_{\alpha}(i)u_{\beta}(j)
    + \frac{e^2}{2}\sum_{i,j}\Pi^0(i,j)h(i)h(j)
    \nonumber\\
 &=&\int d^2 {\vec r}(\mu u^2_{\alpha
  \beta}+\frac{\lambda}{2}(u_{\alpha \beta})^2)
\end{eqnarray}
where $u_{\alpha \beta}$ is the strain tensor and $\lambda,\mu$
are the Lame coefficients, whose explicit expression is (\cite{BNT05})
\begin{equation}\label{Lame_Coeff}
\mu=\frac{\eta(s)}{(4\pi)^{1+s/2}}\frac{e^2}{R^{2+s}}M^{1+s/2} , \
\
\lambda=\frac{\varsigma(s)}{(4\pi)^{1+s/2}}\frac{e^2}{R^{2+s}}M^{1+s/2}
\end{equation}
for $0<s<2$, $\varsigma=\infty$, but this is not important here (it becomes a constraint forcing the
incompressibility of the crystal, but it has been shown that at leading order
in $M$, the relevant elastic constant is the Young Modulus, which remains
finite \cite{BNT05}). It should be recalled that Eq.~\ref{energy_manifold}
involves terms which are not quadratic, so some higher order terms have
been included. Combining Eq.~\ref{energy_manifold} and
Eq.~\ref{Lame_Coeff} it follows
\begin{equation}\label{LargeM}
   \int d^2 {\vec r}(\mu u^2_{\alpha
  \beta}+\frac{\lambda}{2}(u_{\alpha \beta})^2)= M^{1+s/2}
    \frac{e^2}{R^{s+2}} C(s,[u])
\end{equation}
where the $C-$function is defined from
\begin{equation}\label{C_coeff}
 C(s,[u]) =\frac{1}{(4\pi)^{1+s/2}}\int d^2 {\vec r}(\eta(s) u^2_{\alpha
  \beta}+\frac{\varsigma(s)}{2}(u_{\alpha \beta})^2).
\end{equation}
The $C-$function has dimensions of area. Combining Eq.~\ref{LargeM},Eq.~\ref{energy_manifold} and
Eq.~\ref{energy_planar}, the energy Eq.~\ref{Eq_elast} becomes
\begin{eqnarray}\label{energy_total}
E(M)&=&\frac{e^2}{R^s}\left(\frac{M^2}{2^{s}(2-s)}+(\frac{\theta(s)}{2(4\pi)^{s/2}}+\frac{C(s,[u])}{R^2})M^{1+s/2}\right.
\nonumber\\
&+&\left. {\cal O}(M^{s/2}) \right)
\end{eqnarray}
Let us now discuss the dependence of the $C-$coefficient
on the sphere radius $R$. For any configuration $[u]$, we expect
$C(s,[u]) \propto R^2$ and this implies that the energy at order
$M^{1+s/2}$ is increased with
respect to the planar value Eq.~\ref{energy_planar}. If, however, one
could find a configuration $[u]$ such that its growth with $R$ is linear at most, then
it follows that
\begin{equation}\label{min_C}
C(s,u) \propto R a \propto \frac{R^2}{\sqrt{M}}
\end{equation}
where $a$ is the lattice constant and in the last step, the
relation $\frac{\sqrt{3}}{2}a^2=\frac{4 \pi R^2}{M}$ has been used. In
this case, for very large number of particles, the coefficient
$M^{1+s/2}$ in the energy expansion is given by the planar result
and the configuration $[u]$ becomes a minimum of the energy in the limit
of large number of particles.

We now analyze the approximation made in ignoring higher derivative terms in
Eq.~\ref{energy_manifold}. Those terms will consist in higher derivatives of the
strain tensor, which in turn will imply that the elastic constants, equivalent to
the Lame coefficients, contribute to terms growing more slowly than
$M^{s/2+1}$, that is, the ignored terms do not affect the leading term in the
expansion defined by Eq.~\ref{energy_total}.

In generalizing this expansion to other geometries there are several aspects to consider.
First of all, the leading term for long-range potentials scales like $M^2$. The general expression
of the coefficient is
\begin{equation}\label{electrost_fun}
    E^{0}=\frac{e^2}{2}\int d^2x d^2y \rho(x)\frac{1}{|x-y|^s}\rho(y)
\end{equation}
where $\rho(x)$ is the continuum density (The density of particles at scales much larger than a).
In a sphere, $\rho$ is constant and given by $\rho=\frac{M}{4\pi R^2}$, but in a general geometry, the
density follows from minimizing the previous equation
under the constraint $\int d^2x \rho(x)=M$ (For $s=1$ this amounts to solving the
Poisson equation for a fixed density of charges). On a torus, for example, the resulting continuum density
$\rho(x)$ is not constant for $s<2$. That is, the minimization of the leading coefficient for long-range
potentials imposes density variations and invalidates the form of the coefficient ${\cal O}(M^{1+s/2})$
previously discussed. It is possible to generalize the expansion to include density variations but
this will not be done here. On a general geometry, even for short-ranged
potentials, the expansion Eq.~\ref{energy_planar} becomes slightly
more complicated because there maybe more than
one characteristic radius (In a torus, for example, there are two radii) and
the coefficient $C$ will contain a dependence on the dimensionless
parameters that can be constructed from the geometry (for the
torus this is the aspect ratio, the ratio of the two radii), but
the property defining the ground state configuration is still given
by Eq.~\ref{min_C}. The goal is now how to construct those configurations
whose energy grows at most linearly with $R$.

\section{Distribution of dislocations on a triangular wedge}\label{SECT_ansatz}

\begin{figure}[htb]
\begin{center}
\includegraphics[width=3in]{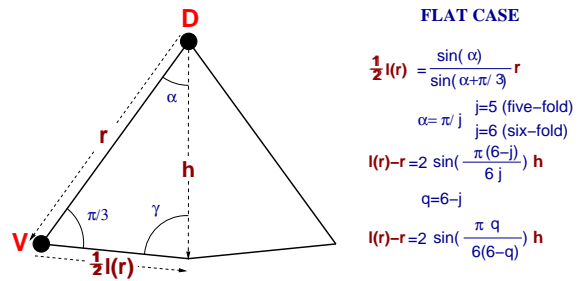}
\caption{(Color online) Triangle considered in the argument}\label{TRI}
\end{center}
\end{figure}

The problem consists now in obtaining the position and location of the
dislocations needed to correct for the variable lattice constant, as discussed
in the introduction. This will be obtained from the following geometric argument.
Let us consider a triangular patch of a surface like the one shown in
figure~\ref{TRI}. At one vertex ($D$ in the figure), we place either a disclination of
charge $q=1$ (a vertex with five nearest-neighbors), $q=0$ (a
regular six fold vertex) or $q=-1$ (a vertex with seven
nearest-neighbors).

If we assume that a disk is formed out of identical triangular wedges,
the angle $2\alpha$ is given by $2\alpha=\frac{2\pi}{6-q_i}$ (
for $q=+1$, $2\alpha=\frac{2\pi}{5}$). We now discuss the necessary conditions
so that the wedge may be triangulated with many equilateral triangles. For that
matter, we now compute by how much
the length of segment $DV$, which is given by $r$, differs from
the length of the segment whose origin is at V and forms 60 degrees
with the segment $DV$ (because we
assume that the triangle whose vertex is at V is equilateral).
The quantity to compute is $l(r)-r$, which from the
purely geometric arguments outlined in fig.~\ref{TRI} for the geometry
of a plane is given by
\begin{equation}\label{geometric}
    l(r)-r=2 \sin(\frac{q_i \pi}{6(6-q_i)})h .
\end{equation}
if $l(r)-r=0$, which happens when $q_i=0$, the wedge can be tiled with
all perfect equilateral triangles, but if the central disclination is
non-zero, this is not possible.
As explained in the introduction, we can
fix this situation by including (or removing) additional rows of particles every time
the equation
\begin{equation}\label{eq_spac}
    l(r)-r=\pm a
\end{equation}
is satisfied, where $a$ is the lattice constant. In physical terms
we are adding a dislocation. In the planar case, the formula above
implies that we add a grain boundary of equally spaced dislocations,
where dislocations within the grain are separated a distance
$D=\frac{a}{2\sin(\frac{q_i \pi}{6(6-q_i)})}$. This result is well
known in metallurgy \cite{HiLo:92}, where grain boundaries of dislocations
in the plane have been extensively investigated .

In an arbitrary geometry, the function $l(r)$ will also depend on
the amount of Gaussian curvature enclosed within the
triangular wedge, and this will lead to a different type of grain
boundaries. The function $l(r)$ is now obtained from differential geometry,
\begin{eqnarray}\label{geometric_geo}
  l(r) &=& \int^{\varphi_a}_{-\frac{\pi}{m}} d\varphi \sqrt{g_{a b}
    (\varphi) V^a(\varphi) V^b(\varphi)} \nonumber \\
    &+& \int^{\frac{\pi}{m}}_{\varphi_a} d\varphi \sqrt{g_{a b}
    (\varphi) V^a(\varphi) V^b(\varphi)}
\end{eqnarray}
where $g_{ab}$ is the metric of the geometry, $V^a$ is the tangent
vector of a geodesic, described by coordinates $\varphi$, which
starts at point $(r,-\frac{\pi}{m})$, which is the equivalent of
point $V$ in figure~\ref{TRI}, and forms an angle of 60 degrees
with the direction defined by the geodesic defined by the radial
distance $r$. Once the function $l(r)$ is known, the location of
the additional dislocations needed to ensure equilateral triangles
will be obtained from the function $l(r)-r$ from
Eq.~\ref{eq_spac}, just as it was done for the planar case. In
this paper, the distance $l(r)$ is computed from a geodesic, but
in general this is not necessarily the case, and examples will be
given when discussing the torus. It should be pointed out that the
predictions that follow from Eq.~\ref{eq_spac} and
Eq.~\ref{geometric} are equivalent to a similar formula provided
in \cite{TrBCN:02} only in the plane, and differ on any other
geometry.

The simplest non-trivial example involving curvature is a
spherical cap. We introduce a new parameter $\theta_M$ that defines the angle
subtended by the triangular wedge. We assume
\begin{equation}\label{sphere_angle}
    \theta_M=L \frac{a}{R}
\end{equation}
where $a$ is the lattice constant, $L$ the total number of particles along the radial
direction $r$ and $R$ the radius of the sphere. We consider a spherical cap both
with a $q=1$ disclination and without any disclination $q=0$ at
its center, each case relevant for large and small aperture angle
$\theta$ respectively.

\begin{figure}[htb]
\begin{center}
\includegraphics[width=3in]{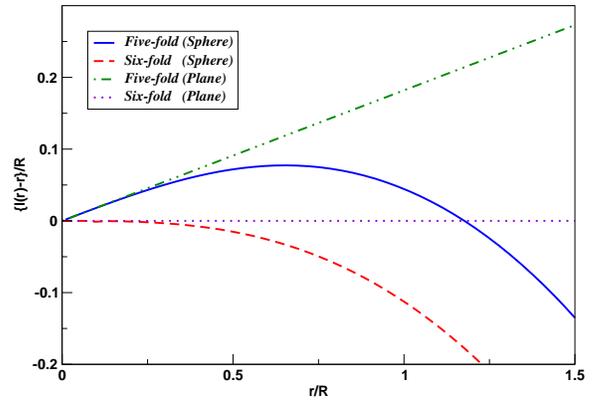}
\caption{(color online)$l(r)-r$ on a sphere for both $q=1,0$ and for a
comparison, the results for the plane are also
shown}\label{Sphere_Plane}
\end{center}
\end{figure}

\begin{figure}[htb]
\begin{center}
\includegraphics[width=3in]{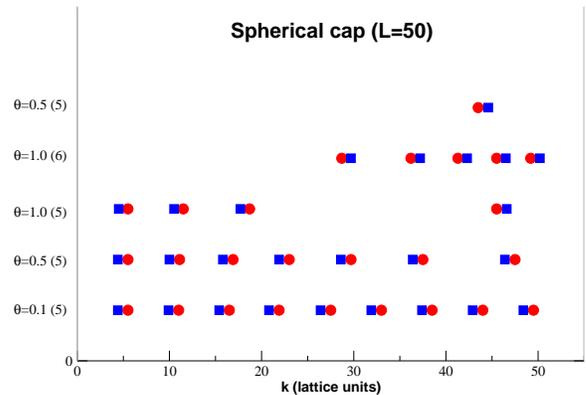}
\caption{(Color Online) Characteristic of grain boundaries of a spherical cap
with different aperture angles with and without a disclination at
the origin. Squares are seven-fold vertices and spheres are five-fold vertices.
This plot corresponds to $L/a=50$.}\label{Sphere_Spacing}
\end{center}
\end{figure}

The different functions $l(r)-r$ are shown in
figure~\ref{Sphere_Plane} for the cases of a five-fold ($q=1$) and
$(q=0)$ disclination and compared to the equivalent results for
the plane. The grain boundaries that follow from the function
above and Eq.~\ref{eq_spac} are shown in
figure~\ref{Sphere_Spacing} for the particular situation $L=50$. In the figure, squares
represent seven-fold and circles represent five-fold vertices, with dislocations being
represented as a five-seven pair. For large aperture angles the last dislocation
has rotated 180 degrees, and this is
a result of the function $l(r)-r$ becoming negative at $r/R$ larger than 1.

\begin{figure}[htb]
\begin{center}
\includegraphics[width=3in]{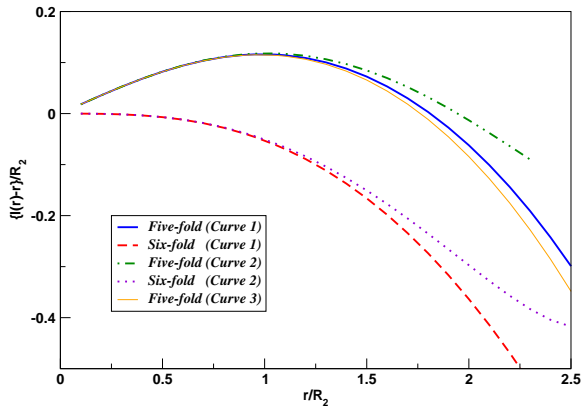}
\caption{(Color online) $l(r)-r$ on a torus at $r=1.2$ for both
$q=1,0$. If the central defect is oriented as shown in
figure~\ref{torus_s} only three functions are needed for
$q=1$(fivefold). In this case $R_1/R_2=1.2$ and
$\beta=0$.}\label{Torus_functions}
\end{center}
\end{figure}

Generalizations to any other geometry are now possible. As an
example, we discuss a wedge of a torus. The critical difference
with the sphere is that the Gaussian curvature in a torus is not
constant, and that implies that the function $l(r)$ is different
for the different wedges of a central defect. For a disk with a five-fold
defect at its center, the five grain boundaries, which are identical for the sphere
become now different. The second difference is that the line joining the
dislocations is not straight, but curved, and depends
on the torus aspect ratio
\begin{equation}\label{torus_ratio}
    r=\frac{R_1}{R_2} \ ,
\end{equation}
where $R_1$ and $R_2$ are the two torus radii. As an example the
functions are shown for an aspect ratio $r=1.2$. In this case, as
shown in figure~\ref{Torus_functions} only three of the functions $l(r)$
become different.

\section{Energy of optimal configurations}\label{SECT_Min}

The previous argument provides the exact location
where dislocations need to be added. Next, we compute the energy
of these configurations. We will consider the
energy for a large, yet finite number of particles. The elastic
energy Eq.~\ref{energy_manifold} can be discretized as
\cite{Trav:03}
\begin{equation}\label{Eqn1}
E_1(\varepsilon,\sigma)
=\frac{\varepsilon}{2}\sum_{(b,c)}(|\vec{r}_{bc}|-a)^2+\sigma
\sum_{<f,d>}(\frac{1}{2}-\cos(\theta_{d f}))^2.
\end{equation}
where $(b,c)$ runs over edges defined by nearest neighbors and angles.
This energy has a very clear
geometric interpretation as providing an energy cost for those triangles
that either are formed of edges whose length is not the lattice constant,
or whose angles are not exactly $60^o$. The coefficients $\varepsilon,\sigma$ are
linearly related to the Lame coefficients \cite{SeNel:88,Trav:03}.

We now evaluate the energy function Eq.~\ref{Eqn1} in the case
$\sigma=0$ for the wedge discussed in the previous section. By
construction, the length of the edges along the radial
distances $r$ are exactly given by $a$, the lattice constant, and
the contribution to the energy of these edges is zero, consistent with
the total radial length of the wedge given by $H=L a$, where $a$ is the
lattice constant and $L$ the total number of edges. The main
contribution to the energy then comes from the vertices along the
direction defined by $l(r)$. At radial distance $r=k a$ there are
$n(k)$ of these vertices, and therefore, the average length of nearest
neighbors along the radial distance $ka$ is $l(ka)/n(k)$. The energy Eq.~\ref{Eqn1}
becomes
\begin{equation}\label{Eqn2}
    E_1(\varepsilon)=\frac{\varepsilon}{2}\sum_{k=1}^L
    \frac{(l(ka)-n(k)a)^2}{n(k)}+ N_d E_{elas} .
\end{equation}
The second term is proportional to the total number of dislocations
$N_{d}$ and takes into account that next to a dislocation the strains
are significant. This energy includes not only the core energy but also the
(exponentially) small distortions arising from the grains. At this point
it is illustrative to show the implications of the
previous formula for a simple and well known situation. We now compute
the energy of a disk consisting of no defects other than a
central disclination at the origin. In that case $l(r)=br$, where
$b$ is given from figure~\ref{TRI}, and since
no additional dislocations are added it is $n(r=ka)=k$. The above formula gives
(with $N_d=0$)
\begin{equation}\label{No_dislocations}
    E_1(\varepsilon)=5\frac{\varepsilon}{2}\int_0^R
    \frac{dr}{r}(l(r)-r)^2=\frac{5 \varepsilon}{4} (ba)^2 L^2
\end{equation}
as expected, the energy of an isolated disclination grows
quadratically with $L^2$. Thus the $C$-coefficient defined by
Eq.~\ref{C_coeff} will grow like $H^2$, and lead to additional contributions
to Eq.~\ref{energy_total} (here $H=La$ plays the role of $R$ in the sphere).
All these results are well known
\cite{SeNel:88,Trav:03} but it is of interest to show how they are
recovered within the previous approach. It is now easy to show
that the energy of the configurations that satisfy
Eq.~\ref{eq_spac} grow more slowly than $R$. We have,
\begin{eqnarray}\label{Eqn2_pro}
  E_1(\varepsilon)&=&\frac{\varepsilon}{2}\sum_{k=1}^L
    \frac{(l(ka)-n(k)a)^2}{n(k)}\nonumber\\
    &<& \frac{\varepsilon}{2}\sum_{k=1}^L
    \frac{1}{n(k)} a^2 \sim log(L)
\end{eqnarray}
where the last step follows because by construction $l(r)-n(r)a$
never exceeds one. Within the same assumptions, it is very simple to repeat the argument for the
second term in Eq.~\ref{Eqn1} and the same logarithmic behavior is found, so the statement
holds for the entire range of elastic constants. As already mentioned, distortions near dislocations
are not entirely negligible, and this leads to a term that grows
linearly with $L$, .
\begin{equation}\label{Eqn3_pro}
E_1(\varepsilon)=N_d E_{core} \sim H a = L a^2.
\end{equation}

\begin{figure}[htb]
\begin{center}
\includegraphics[width=3in]{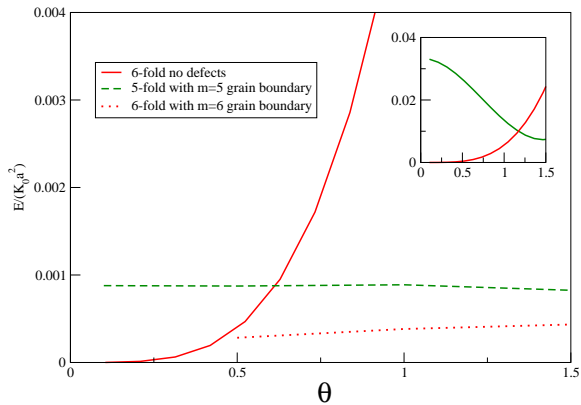}
\caption{(Color Online) Energy of the spherical cap (L=50) with
and without defects. The contribution proportional to the number
of dislocations is not shown.}\label{Sphere_theta}
\end{center}
\end{figure}

As an example of the previous considerations, we discuss the
spherical cap. We determine best distributions as a function of
$\theta_M$, the subtended angle, and $L$ the radial number of
particles. As $\theta_M \rightarrow 0$, the spherical cap becomes
a plane and the ground state should approach the energy of a perfect
planar lattice. As $\theta_M$ is increased, Gaussian curvature effects
become important, and a disclination at the top of the cap lowers the
energy as shown in the inset of fig.~\ref{Sphere_theta}. Upon including
dislocations the energy decreases enormously and remains essentially
independent with the subtended angle in both cases.

The previous considerations do show that the geometrical ansatz
does provide a minimum energy configurations, albeit a degenerate one. For the spherical
cap, for example, the minimum energy can be achieved either by a
plus disclination or without any disclinations at all, provided
that the appropriate grain boundaries, as defined by Eq.~\ref{eq_spac} are
used. The next question is therefore, which one of these
minima is the actual ground state of the system. This involves considering
the sub-leading coefficients in the expansion. From
Eq.~\ref{Eqn2_pro} this is given by the configuration with
the lowest possible number of defects. This still does not completely solve the
problem. One can consider several grain boundary on each triangular wedge where
the separations of dislocations within the grain is larger, thus keeping the
total number of defects constant. For the planar case, this question was
investigated in \cite{Trav:03} where it was concluded that grain boundaries with the smallest
spacing within dislocations are favored. Further numerical investigations will
hopefully provide more evidence on this point.

\section{Solutions for the Sphere and the Torus}\label{SECT_patch}

\subsection{The Sphere}

There are several topological inequivalent triangulations of a
sphere with icosahedral symmetry, and are labelled by two integers
$(n,m)$. We just describe here in detail the solutions to the
cases $(n,n)$ and $(n,0)$, but solutions of the form $(n,m)$ may
be constructed along the same lines. The problem consists now in dividing
the sphere into triangular wedges such that can be consistently
joined back together after the additional dislocations necessary to
relieve the geometric frustration have been included.

\begin{figure}[htb]
\begin{center}
\includegraphics[width=2in]{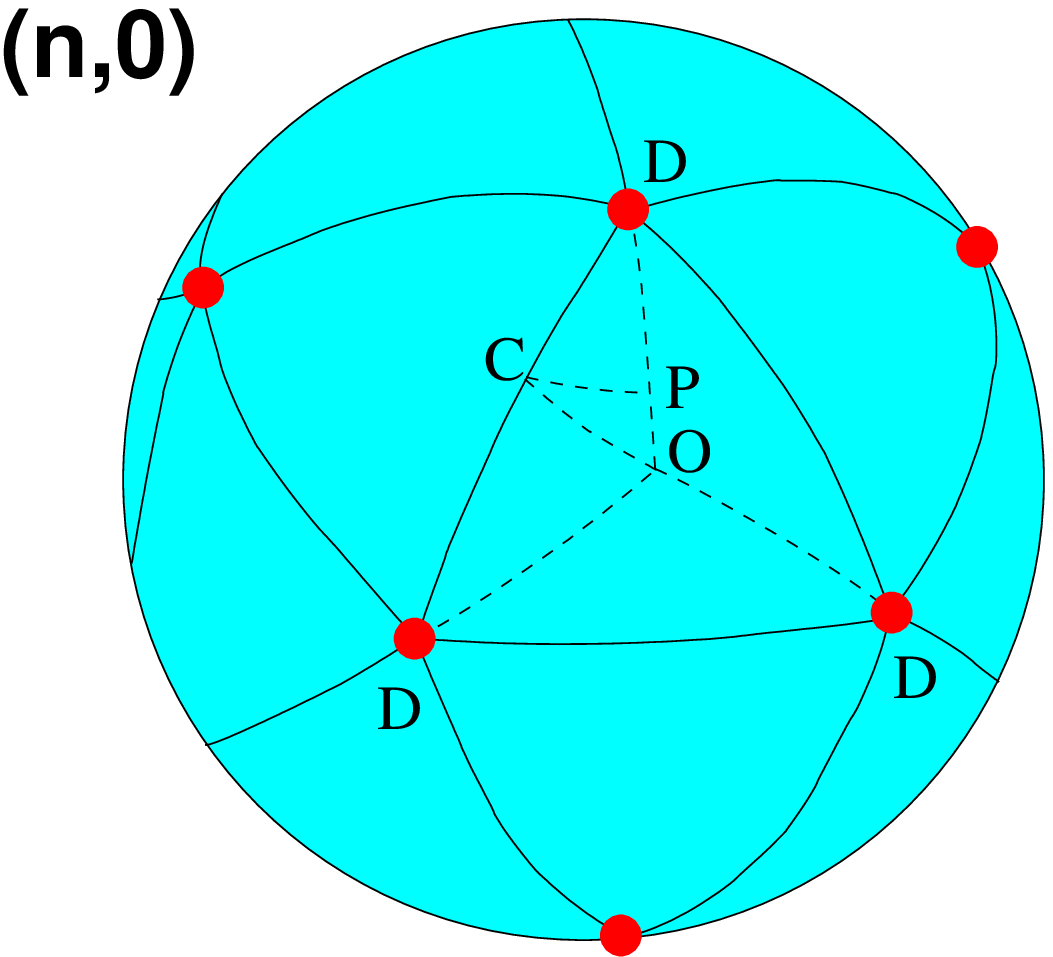}
\includegraphics[width=2in]{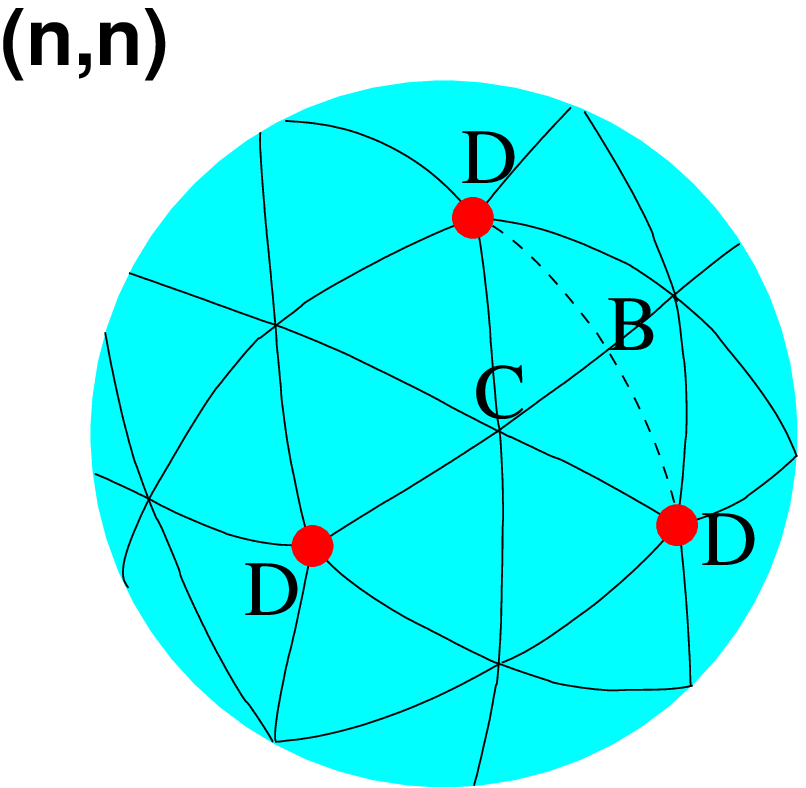}
\caption{(Color Online)Triangular wedges used to solve the $(n,n)$
and the $(n,0)$ configurations. Dashed lines follow the directions
of the grain boundaries. Filled circles represent + disclinations,
also marked with D.}\label{Sphere_nn_n0}
\end{center}
\end{figure}

The $(n,n)$ configuration can be divided into 20 triangular wedges
like the one shown in fig.~\ref{Sphere_nn_n0} (2 triangular wedges are
shown). From the results described for spherical wedges, the
dislocations follow the line AB, and the spacing is predicted
from the function fig.~\ref{Sphere_Plane} for an aperture angle
$\theta_M\equiv\theta_V=\arcsin(2\sqrt{1/2-1/{(2\sqrt5)}}/\sqrt
3)\approx 37^o$. The fact that the angle ABC is 90 degrees
and the angle ACB is 60 degrees, which can be checked by using formulas
in spherical trigonometry, ensures that neighboring triangular wedges
are perfectly joined and the complete consistency of the solution.

The triangular wedges for the $(n,0)$ are shown in
fig.~\ref{Sphere_nn_n0}. A natural patch is defined by DCO, but it should be
noted that CO does not define a row of particles of the crystal. In this case,
the patches extend beyond point C, and overlap slightly. The dislocations
follow the lines DO and the spacing is predicted from the results from
the function fig.~\ref{Sphere_Plane} similarly as in the $(n,n)$ case. Here, however,
the consistency is a little more difficult to check, because in the region
defined by such triangles such as COP, which spans aperture angles between
$(\theta_G,\theta_V)$, where
$\theta_M\equiv\theta_G=\arccos(1/\sqrt(5))/2\approx 32^o$, the different
patches overlap. The consistency here is due to the sum of the Burgers vectors
of the three grain boundaries that approach point O adding to zero, which ensures
that all rows of particles added terminate within the patches.

\subsection{The Torus}

\begin{figure}[htb]
\begin{center}
\includegraphics[width=3in]{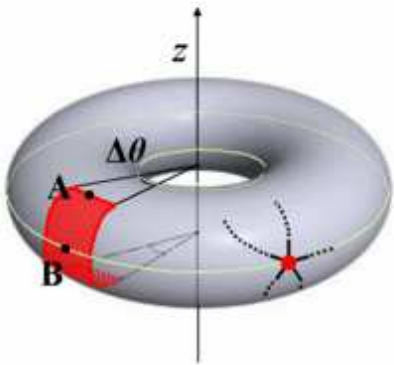}
\caption{(Color Online)Representation of the torus, with the definition of
the subtended angle. The two circles are geodesics of maximum and
minimum curvature. A fivefold defect on the positive curvature of the
torus has been shown.}\label{torus_s}
\end{center}
\end{figure}

The torus is depicted in fig.~\ref{torus_s}. There are two
radii curvature, $R_1,R_2$, and the aspect ratio is defined by
$r=\frac{R_1}{R_2}>1$. As already discussed the only situation we discuss concerns
short-ranged potentials. The Gaussian
curvature of the torus depends only on the coordinate $\varphi$, and it is
\begin{equation}\label{torus_curv}
    K=\frac{\cos(\varphi)}{R_2^2(r+\cos(\varphi))} .
\end{equation}
Since Gaussian curvature attracts like sign disclinations
\cite{BNTr:00}, we assume here that the ground state of the torus
contains 12 positive disclinations located along the geodesic of maximum
curvature (outer curve) and 12 negative disclinations along the geodesic
of minimum curvature (inner curve), and several grain boundaries of dislocations.
Another possibility would consist of a ground state consisting of grain
boundaries of dislocations only. It is very likely that for a thin torus $ r>>1$
disclinations may not be favored because nearest-neighbor
disclinations are so far away, that radial grain boundaries may not efficiently
screen the strains.

\begin{figure}[htb]
\begin{center}
\includegraphics[width=3in]{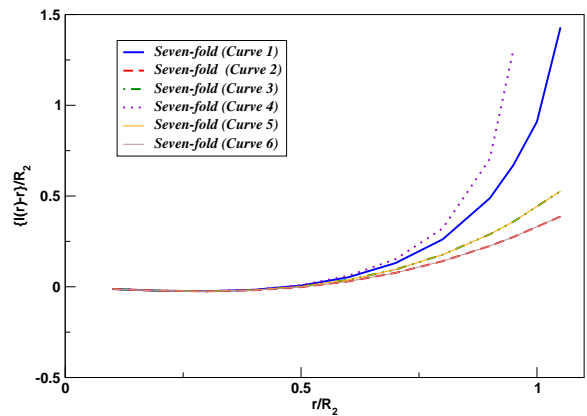}
\caption{(Color Online)$l(r)-r$ on a torus at $r=1.2$ for both
$q=1$. The different functions are shown. ($R_1/R_2=1.2$ and
$\beta=0$)}\label{Torus_functions_neg}
\end{center}
\end{figure}

If the ground state contains disclinations, the grain
boundaries that appear following the five-fold disclinations have already been computed in
figure~\ref{Torus_functions}. There is a similar function for the seven-fold disclinations, which
are located along the interior circle in the figure. The function predicting its spacing
is given in figure~\ref{Torus_functions_neg}.

The situation were the ground state of the torus consists of grain boundaries of dislocations
only is qualitative different, because in that case, the grain boundaries are not radial, but
follow the directions defined by the azimuthal angle, the
curve $AB$ in figure~\ref{torus_s}, and a similar curve with opposite oriented dislocations
on the inner side of the torus. The particles are located in rings defined by the condition
$\psi=constant$, and therefore, only $\psi=0,\pi$ are geodesics, which correspond to the
two circles in fig.~\ref{Torus_functions_neg}. Similarly as in the sphere, there are other
torus triangulations that maybe constructed from twisting and rejoining the lattice either along
the $varphi$ or the $\theta$ direction, but the discussion of the grain boundaries in this case
is beyond the scope of this paper.

\section{Conclusions}\label{SECT_conclusions}

We now summarize some of the main predictions that follow from the results presented in this paper that can
be verified from either numerical simulations or experimental results:
\begin{enumerate}
    \item The ground state consists of grain boundaries of dislocations whose exact
    position and orientation is predicted from Eq.~\ref{eq_spac}. Explicit examples have
    been provided for both the sphere and the torus.
    \item The Energy of the ground state tends to a universal value for very large
    number of particles, which is independent of the geometry and given by Eq.~\ref{energy_planar}.
    (For long-range potentials, there is an additional term, which grows quadratically with the
    number of particles).
    \item For potentials $V(r)=\frac{e^2}{r^{s}}$, the sub-leading corrections to the
    ground state are of order $M^{(s+1)/2}$ in the number of particles.
    \item The ground state is degenerate at leading order, as there are some free
    parameters characterizing the grain boundaries. This degeneracy is removed at sub-leading
    order by those configurations with the minimum number of dislocations and the lowest possible
    number of grain boundaries.
\end{enumerate}

There are several ways to verify these predictions. The main difficulty with numerical
minimizations is that it is very difficult to prepare an initial configuration that will relax
to a previously selected distribution of defects. For the Thomson problem, however, the
ring-removal technique developed by Toomre \cite{Toomre:97} seems to generate the type of grain boundaries
that this paper predicts as the ground state. In fact, in \cite{PGMoo:99}, it was shown that these grain
boundaries significantly lower the energy, bringing the icosadeltahedral configurations closer to
the planar limit Eq.~\ref{energy_planar}. It was further speculated that the true ground state
could be achieved by successive applications of the technique, but this statement was not substantiated
because the actual rings that needed to be removed were not known. The function in figure~\ref{Sphere_Plane}
together with Eq.~\ref{eq_spac} does provide the location of the dislocations
and therefore the actual rings that need to be removed, so the ring removal technique appears as a very
promising practical tool to verify predictions 1 to 3 for the Thomson problem. It would also be of great
interest to check how this predictions compare with ground states for short-ranged potentials, where the results
of this paper are more rigorously justified.

Evidence on the validity of statements 1-4 has also been provided from numerical and analytical
results in \cite{Trav:03}, where by using a discretized version of elasticity theory, it is shown
that 1-4 are verified for a plane with the constraint of total disclination charge equal to $\pm1$.
The methods presented, however, are far general and it is expected that the extension of the
results for any geometry will soon follow.

Statement 2) is in agreement with recent rigorous results
proven for potentials $s>2$ \cite{HarSa:05,HarSa:04}. These results have greater
generality, since they apply to dimensions other than 2. Preliminary results for
the torus with short-ranged potentials have recently become available
\cite{Wommer:05,BCMi:05} and show density variations for $s<2$ and ground states for $s>2$
with and without disclinations, but the present state of the simulations do not allow to draw
a more quantitative analysis. Experimental evidence can also be used to prove the validity
of statement 1. In \cite{Science:03} it was shown that PMMA beads assemble on an spherical oil-water
interface forming spherical crystals, which can then
be imaged by confocal microscopy. The current experiments show that next to five-fold defects additional
dislocations arise, but there are only two of them, instead of the five predicted by this paper (Interestingly
this is the number favored in the flat case \cite{Trav:03}). Furthermore,
the dislocations within the grain show a constant spacing. This is possibly due to the fact that the total
number of particles is still too small for the asymptotic results of this paper to apply. Future experiments
with larger particle aggregation numbers should settle this issue.

An important point that has not been addressed in this paper is
the critical value of $M$ at which the asymptotic solution
proposed will apply. A numerical verification of the predictions
of this paper is currently under way for short-ranged potentials
and preliminary results indicate that for the sphere, this number
is of the order of five thousand particles at the very least. We
hope to report more on this in the near future.

In applying the formulas derived to a real situation, the
dislocations must be located at points in the crystal, and
accordingly, the spacing of the dislocations must always be an
integer. To apply the previous formulas, the location of the
dislocations must therefore be rounded to the closer integer. The
resulting orientations of the dislocations may also not be
entirely consistent with the location of the crystallographic axis
of the crystal, and this may require, for example, placing
dislocations separated by an odd number of lattice constants
\cite{Trav:03}. Further numerical work will hopefully clarify this
more technical points.

There are a certain number of issues that this paper has not addressed. First of all, the
ground states discussed for both the sphere and the torus rigorously apply only for some "magic"
number of particles $M$. We expect that for large number of particles outside this magic numbers,
the ground state solutions should not be that different from the ones proposed in this paper,
since the range of $M$ in between magic numbers is small compared with $M$ itself. We hope that
future numerical work will address this issue.

In this paper, it has been assumed that the potential is isotropic. If the potential is not isotropic the ground
state on a plane may not be a triangular lattice, and the present arguments need to be modified.
For very large number of particles, it should be expected that locally, the triangulation will be
very close to the flat case, so similarly as in the
isotropic case, we expect that additional dislocations will be required
to fix the frustration induced by the geometric constraints.

There has been recent interest in understanding similar problems
as the one discussed in this paper where liquid crystalline order
is discussed on a non-zero geometry. Examples include hexatic
\cite{BNTH:04,ViTu:04,ViNel:04}, nematic \cite{Nelson:02}, or
smectic blue phases \cite{DDKa:03}. The results discussed in this
paper, where the energy is regarded as an expansion in $M$ and the
defects correct for the geometric frustration are completely
general and may be generalized to this cases as well. We hope to
report more in the near future.

{\bf Acknowledgments}

This paper has arisen from many discussions with Mark Bowick and David Nelson.
Discussions with Angelo Cacciuto,Slava Chushak, Doug Hardin and Edward Saff
are also acknowledged. This work has been supported by
NSF grant DMR-0426597 and by Iowa State Start-up funds.


\begin{thebibliography}{10}
\bibitem{com}
We are assuming isotropic potentials. If the potential is not isotropic,
in that case the ground state in the plane may be different than the
triangular lattice.
\end{thebibliography}
\end{document}